\documentclass[aps,pra,preprint,groupedaddress,amsmath,amssymb,showpacs]{revtex4}
\usepackage{graphicx,amsmath,amssymb,epsfig}
\usepackage{dcolumn}% Align table columns on decimal point
\usepackage{bm}
\usepackage{color}
\begin{document}
%\preprint{}

\title{Ultraslow Optical Solitons and Their Storage and Retrieval in an Ultracold
Ladder-Type Atomic System}

\author{Yang Chen}
\affiliation{State Key Laboratory of Precision Spectroscopy and Department of Physics,
East China Normal University, Shanghai 200062, China}

\author{Zhengyang Bai}
\affiliation{State Key Laboratory of Precision Spectroscopy and Department of Physics,
East China Normal University, Shanghai 200062, China}

\author{Guoxiang Huang}
\email[Email: ] {gxhuang@phy.ecnu.edu.cn}
\affiliation{State Key
Laboratory of Precision Spectroscopy and Department of Physics, East
China Normal University, Shanghai 200062, China}

\begin{abstract}

We propose a scheme to obtain stable nonlinear optical pulses and realize their storage and retrieval
in an ultracold ladder-type three-level atomic gas via electromagnetically induced transparency. Based on Maxwell-Bloch equations we derive a nonlinear equation
governing the evolution of probe field envelope, and show that
optical solitons with ultraslow propagating velocity and extremely low
generation power can be created in the system. Furthermore,
we demonstrate that such ultraslow optical solitons
can be stored and retrieved by switching off and on a control field.
Due to the balance between dispersion and nonlinearity,
the ultraslow optical solitons are robust during propagation,
and hence their storage and retrieval are more desirable than that of
linear optical pulses. This raises the possibility of realizing
the storage and retrieval of light and quantum information
by using solitonic pulses.


\end{abstract}
\maketitle

\section{INTRODUCTION}{\label{sec1}}

%{\color{blue} hh}

In recent years, much effort has been paid to the study of electromagnetically induced
transparency (EIT), a typical quantum interference effect occurring in resonant multi-level
atomic systems. The light propagation in EIT systems possesses many striking features,
including substantial suppression of optical absorption, significant reduction of group
velocity, giant enhancement of Kerr nonlinearity, and so on.
Based on these features, many applications of
nonlinear optical processes at weak light level can be realized~\cite{FIM}.

One of important applications of EIT is light storage and retrieval,
which can be explained by the concept of dark state polariton~\cite{Fleischhauer2000},
a combination of atomic coherence and probe pulse. The dark state polariton prominently
shows atomic character when a control field is switched off and
the optical character when the control field is switched on. The
storage and retrieval of probe pulses based on EIT have been
verified in many experiments~\cite{LDBH,phillip2001,Longd,schnorr2009,Zhao,zhang2009,Rad,Dudin10,
Yangf,PFMWL,dudin2013,Dai,heinze2013}.

However, up to now most of previous works on light storage and retrieval based on EIT are
carried out in $\Lambda$-type three-level atomic systems. In addition, the probe pulse used
is very weak and hence the system works in linear regime, except for the
numerical study presented in Ref.~\cite{dey2003}.
It is known that linear probe pulses suffer a spreading and attenuation due to
the existence of dispersion, which may result in a serious distortion for retrieved
pulses. For practical applications, it is desirable to obtain optical pulses that
are robust during storage and retrieval.

In this article, we propose a scheme to produce stable weak nonlinear optical pulses and
realize their robust storage and retrieval. The system we consider is
an ultracold atomic gas with a ladder-type three-level configuration working under
EIT condition. Starting from Maxwell-Bloch (MB) equations we derive a nonlinear
Schr\"{o}dinger (NLS) equation governing the evolution of probe-field envelope,
and show that optical solitons with ultraslow propagating velocity and extremely low
generation power can be created in the system. Furthermore,
we demonstrate that such ultraslow optical solitons
can be stored and retrieved by switching off and on a control field.
Due to the balance between dispersion and nonlinearity,
the ultraslow optical solitons are robust during propagation,
and hence their storage and retrieval are more desirable than that of
linear optical pulses.

Before preceding, we note that, on the one hand, recently much attention has focused on ultracold Rydberg
atoms~\cite{saf,pri0} due to their intriguing properties useful for quantum information processing and nonlinear optical processes; on the other hand, ultraslow optical solitons
via EIT have been predicted for $\Lambda$-type three-level atoms~~\cite{deng2004,huang2005}.
In a recent work Maxwell {\it et al.}  reported the storage
of weak (i.e. linear) probe pulses in a ladder-type system using ultracold Rydberg atoms~\cite{Maxwell2013}.
However, to the best of our knowledge, till now there has been no report on the storage and retrieval of
ultraslow optical solitons in ladder-type atomic systems.
Our theoretical results given here raise the possibility of realizing the storage and retrieval
of light and quantum information by using nonlinear solitonic pulses. Experimentally, it is hopeful to
employ low-density ultracold Rydberg atoms, where the Rydberg state has a very long
lifetime~\cite{Saf,Pritch}, to realize the storage and retrieval of the
ultraslow optical solitons predicted in our work.

The article is arranged as follows. In Sec.~\ref{Sec2}, the physical model under study is described. In Sec.~\ref{sec3}, a derivation of NLS equation controlling the evolution of probe field envelope is given, and ultraslow optical soliton solutions and their interaction are presented.  In Sec.~\ref{sec4},  the storage and retrieval of the ultraslow optical solitons are investigated in detail. Finally, the last section contains a summary of the main results of our work.

%%%%%%%%%%%%%%%%%%%%%%%%%%%%%%%%%%%%%%%%%%%%%%%%%%%%%%%%%%%%%%%%%%%%%%%%%%%%%%%%%%%%%%%%%%%%%
\section{MODEL}{\label{Sec2}}

\begin{figure}
\centering
\includegraphics[scale=0.7]{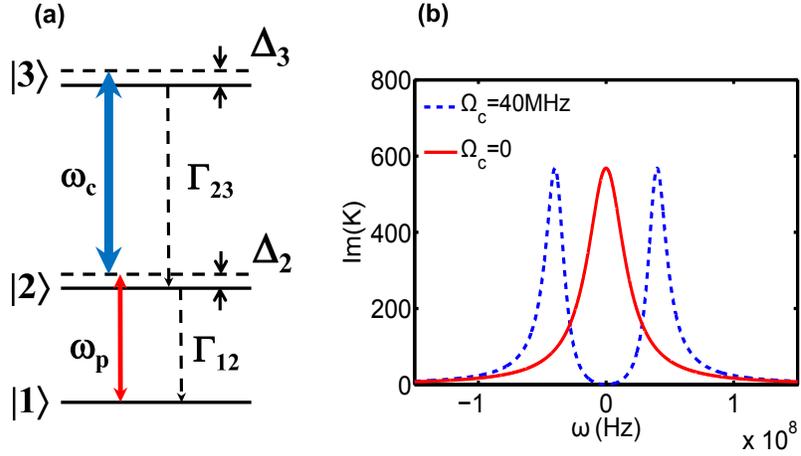}
\caption{(Color online) (a) Energy-level diagram and excitation scheme
of the three-level ladder-type atoms, in which the states $|2\rangle$ and $|1\rangle$ are
coupled by the probe field with angular frequency $\omega_p$, and the states $|3\rangle$
and $|2\rangle$ are coupled by the control field with angular frequency $\omega_c$. $\Delta_{2}$
and $\Delta_{3}$ are respectively one- and two-photon detunings; $\Gamma_{13}$ ($\Gamma_{23}$) is
the spontaneous emission decay rate from  $|3\rangle $ to $|1\rangle$ ($|3\rangle$ to $|2\rangle$).
(b) The absorption spectrum Im($K$) of the probe field as a function of $\omega$.
The solid and dashed lines correspond respectively to $\Omega_c$ =0 (no EIT) and
$\Omega_c$=40 MHz (EIT).} \label{newmodel}
\label{fig1}
\end{figure}

We consider a life-broadened three-level atomic system with a ladder-type
level configuration (Fig.~\ref{fig1}(a)), where $|1\rangle$, $|2\rangle$, and $|3\rangle$
are ground, intermediate, and upper states, respectively.
Especially, the state  $|3\rangle$ can be taken as a Rydberg state that has
a very long lifetime. We assume the atoms work in an ultracold (e.g. $\mu$K)
environment so that their
center-of-mass motion can be ignored. A probe field of the center angular frequency $\omega_p$ couples to the transition $|1\rangle\rightarrow|2\rangle$ and a control field of center angular
frequency $\omega_c$ couples to the transition $|2\rangle\rightarrow|3\rangle$, respectively.

%
%%%%%%%%%%%%%%%%%%%%%%%%%%%%%%%%%%%%%%%%%%%%%%%%%%%%%%
%%%%%%%%%%%%%%%%%%%%%%%%%%%%%%%%%%%%%%%%%%%%%%%%%%%%%%
%

For simplicity, we assume both the probe and the control fields propagate along the $z$ direction. Then the electric field  of the system can be expressed as $\mathbf{E}$=$\sum_{l=p,c}\mathbf{e}_l{\cal E}_l{\rm exp}[i(k_lz-\omega_lt)]+{\rm c.c.}$. Here $\mathbf{e}_p$ and $\mathbf{e}_c$  (${\cal E}_p$ and ${\cal E}_c$) are respectively the polarization unit vectors (envelopes) of the probe and control field; $k_p=\omega_p/c$ and $k_c=\omega_c/c$ are respectively the wavenumbers of the probe and control fields before entering the medium.

Under electric-dipole and rotating-wave approximations, the equations of the motion for the density
matrix elements in interaction picture read~\cite{julio}
\begin{subequations} \label{bloch}
\begin{eqnarray}
&&\frac{\partial}{\partial
t}\sigma_{11}=i\Omega_p^{*}\sigma_{21}-i\Omega_p\sigma_{21}^{*}
+\Gamma_{12}\sigma_{22},\label{bloch1} \\
&&\frac{\partial}{\partial
t}\sigma_{22}=i\Omega_p\sigma_{21}^{*}+i\Omega_c^{*}\sigma_{32}-i\Omega_p^{*}\sigma_{21}-i\Omega_c\sigma_{32}^{*}
-\Gamma_{12}\sigma_{22}+\Gamma_{23}\sigma_{33},\label{bloch2}\\
&&\frac{\partial}{\partial
t}\sigma_{33}=i\Omega_c\sigma_{32}^{*}-i\Omega_c^{*}\sigma_{32}-\Gamma_3\sigma_{33},\label{bloch3}\\
&&\frac{\partial}{\partial
t}\sigma_{21}=id_{21}\sigma_{21}+i\Omega_p(\sigma_{11}-\sigma_{22})+i\Omega_c^{*}\sigma_{31},\label{bloch4}\\
&&\frac{\partial}{\partial
t}\sigma_{31}=id_{31}\sigma_{31}-i\Omega_p\sigma_{32}+i\Omega_c\sigma_{21},\label{bloch5}\\
&&\frac{\partial}{\partial
t}\sigma_{32}=id_{32}\sigma_{32}+i\Omega_c(\sigma_{22}-\sigma_{33})-i\Omega_p^{*}\sigma_{31},\label{bloch6}
\end{eqnarray}
\end{subequations}
where $\Omega_p=(\mathbf{e}_p\cdot\mathbf{p}_{21}){\cal
E}_p/\hbar$ and $\Omega_c=(\mathbf{e}_c\cdot\mathbf{p}_{32}){\cal
E}_c/\hbar$ are
respectively the half Rabi frequencies of the probe and the control fields,
with $\mathbf{p}_{ij}$ the electric dipole matrix element associated
with the transition between $|j\rangle$ and $|i\rangle$. In Eq.~(\ref{bloch}),
$d_{21}=\Delta _{2}+i\gamma _{21}$, $d_{31}=\Delta
_{3}+i\gamma _{31}$, and $d_{32}=(\Delta _{3}-\Delta _{2})+i\gamma
_{32}$, with $\Delta_2=\omega_p-(\omega_2-\omega_1)$ and
$\Delta_3=\omega_p+\omega_c-(\omega_3-\omega_1) $  respectively the one-photon
and two-photon detunings;
$\gamma_{ij}=(\Gamma_i+\Gamma_j)/2+\gamma_{ij}^{\rm col}$,
$\Gamma _{j}=\sum_{i<j}\Gamma _{ij}$, with $\Gamma _{ij}$ denoting
the spontaneous emission decay rate from $|j\rangle$ to $|i\rangle
$ and $\gamma _{ij}^{\rm col}$ denoting the dephasing rate
between the states $|i\rangle $ and $|j\rangle $~\cite{Boyd}.

The evolution of the electric field is governed by the Maxwell
equation, which under a slowly varying envelope approximation yields~\cite{li08}
\begin{subequations} \label{maxwell}
\begin{eqnarray}
&& \label{Max11} i\left(\frac{\partial}{\partial z}
+\frac{1}{c}\frac{\partial}{\partial t}\right)\Omega_p +\kappa_{12}\sigma_{21}=0,\\
&& \label{Max12} i\left(\frac{\partial}{\partial z}
+\frac{1}{c}\frac{\partial}{\partial t}\right)\Omega_c +\kappa_{23}\sigma_{32}=0,
\end{eqnarray}
\end{subequations}
where
$\kappa_{12}={\cal N}_a\omega_{p}|\mathbf{p}_{12}|^2/(2\varepsilon_0\rm c\hbar)$ and $\kappa_{23}={\cal N}_a\omega_{c}|\mathbf{p}_{23}|^2/(2\varepsilon_0\rm c\hbar)$, with ${\cal{N}}_a$ the atomic density.
Note that for simplicity we have assumed both the probe and control fields have large beam radius in both
$x$ and $y$ directions so that the diffraction effect representing by the term
$(\partial^2/\partial x^2+\partial^2/\partial y^2)\Omega_{p,c}$ can be neglected.

%%%%%%%%%%%%%%%%%%%%%%%%%%%%%%%%%%%%%%%%%%%%%%%%%%%%%%%%%%%%%%%%%%%%%%%%%%%%%%%%%%%
\section{Ultraslow optical solitons}\label{sec3}

\subsection{Nonlinear envelope equation}

We first consider the formation and propagation of ultraslow optical solitons in the system.
We assume the probe field is weakly nonlinear and pulsed with time duration $\tau_0$; the control field is a continuous-wave (i.e. its  time duration is much larger than  $\tau_0$) and  strong enough so that its depletion
can be neglected during propagation, which means  $\Omega_c$ can be taken as a constant and hence Eq.~(\ref{Max12}) can be disregarded.

In order to derive the nonlinear envelope equation of the probe field, we make the asymptotic expansion
$\sigma_{jl}=\sigma_{jl}^{(0)}+\epsilon \sigma_{jl}^{(1)}+\epsilon^2 \sigma_{jl}^{(2)}+\epsilon^3 \sigma_{jl}^{(3)}
+\cdots$, $\Omega_{p}=\epsilon\Omega_p^{(1)}+\epsilon^2\Omega_p^{(2)}+\epsilon^3\Omega_p^{(3)}+\cdots$,
with $\sigma_{jl}^{(0)}=\delta_{j1}\delta_{l1}$ and $\epsilon$ a small parameter characterizing the amplitude
of $\Omega_p$. To obtain a divergence-free expansion, all quantities on the right hand side of the
expansion are considered as functions of the multiscale variables $z_l=\epsilon^{l} z$  $(l=0,1,2)$, and $t_l=\epsilon^{l} t$ $(l=0,1)$.

Substituting the above expansion to the MB Eqs.~(\ref{bloch}) and (\ref{Max11}) and comparing the
the coefficients of $\epsilon^l$ $(l=1,2,3...)$, we obtain a set of linear but inhomogeneous equations which can be solved order by order. At the first order, we obtain the solution
\begin{subequations}\label{firstorder}
\begin{eqnarray}
&& \label{FO1} \Omega_p^{(1)}=F\,e^{i\theta}, \\
&& \label{FO2} \sigma_{21}^{(1)}=\frac{\omega+d_{31}}{D(\omega)}F
e^{i\theta}, \\
&& \label{FO3} \sigma_{31}^{(1)}=-\frac{\Omega_c}{D(\omega)}F
e^{i\theta},
\end{eqnarray}
\end{subequations}
with other $\sigma_{jl}^{(1)}$=0. In the above expressions, $D(\omega)=|\Omega_c|^2-(\omega+d_{21})(\omega+d_{31})$,
$\theta=K(\omega)z_0-\omega t_0$~\cite{note0} with $F$ the envelope function of the slow variables $z_l$, $z_2$, and
$t_1$; $K(\omega)$ is the linear dispersion relation of the system, given by
\begin{equation} \label{dispersion}
K(\omega)=\frac{\omega}{c}+\frac{\kappa_{12}(\omega+d_{31})}{D(\omega)}.
\end{equation}
Fig.~\ref{newmodel}(b) shows Im($K$), i.e. the imaginary part of $K$, as a function of $\omega$.
When plotting the figure, we have chosen ultracold Rydberg atoms with the levels in Fig.~\ref{fig1}(a)
and the system parameters as~\cite{Maxwell2013,adams2010}:
\begin{eqnarray}\label{param}
& & \label{param1}|1\rangle=|5s^2 S_{1/2},F=2\rangle,\,\,\,\,
     |2\rangle=|5p^2 P_{3/2},F=3\rangle,\,\,\,\, |3\rangle=|60s^2S_{1/2},\rangle.\\
& &  \label{param2}\Gamma_{12}/2\pi=6\,\,{\rm MHz}, \,\,\,\,\Gamma_{23}/2\pi=3200\,\, {\rm Hz},\,\,\,\,
    \gamma_{21}\approx 18.8\,\, {\rm MHz}, \,\,\,\, \gamma_{31}\approx 1000\,\, {\rm Hz}.
\end{eqnarray}
In addition, we assume ${\cal N}_a\approx1.79\times10^{11}$ $\rm cm^{-3}$, then $\kappa_{12}$ takes the value $1.0\times 10^{10}$ $ {\rm cm}^{-1}{\rm s}^{-1}$. The solid and dashed line in Fig.~\ref{newmodel}(b) correspond, respectively, to the absence ($\Omega_c=0$) and the presence ($\Omega_c=40$ $ {\rm MHz}$) of the control field. We see that in the absence of $\Omega_c$ the probe field has a large absorption (the solid line in Fig.~\ref{newmodel}(b)\,) (i.e. no EIT); however, in the presence of $\Omega_c$ a transparency window is opened in Im($K$) (the dashed line in Fig.~\ref{newmodel}(b)\,), and hence the probe pulse can propagate in the resonant atomic system with negligible absorption (i.e. EIT). The openness of the EIT  transparency window is due to the quantum interference effect induced by the control field.

At the second order, a divergence-free condition requires
\begin{equation}\label{EQ1}
\frac{\partial F}{\partial z_1}+\frac{1}{V_g}\frac{\partial F}{\partial t_1}=0,
\end{equation}
with $V_g=(\partial K/\partial \omega)^{-1}$ being the group velocity of $F$.
The second-order solution reads $\sigma_{21}^{(2)}= A_{21}^{(2)}(\partial F/\partial
t_1)e^{i\theta}$,  $\sigma_{31}^{(2)}=A_{31}^{(2)}(\partial F/\partial t_1) e^{i\theta}$, $\sigma_{11}^{(2)}=A_{11}^{(2)}|F|^2e^{-2\bar{\alpha}z_2}$,
$\sigma_{33}^{(2)}=A_{33}^{(2)}|F|^2e^{-2\bar{\alpha}z_2}$,
$\sigma_{32}^{(2)}=A_{32}^{(2)}|F|^2e^{-2\bar{\alpha}z_2}$  with
\begin{subequations} \label{secondorder}
\begin{eqnarray}
&& A_{21}^{(2)}=\frac{i}{\kappa_{12}}\left(\frac{1}{V_{{\rm
g}}}-\frac{1}{c}\right),\\
&& A_{31}^{(2)}=\frac{i}{\Omega_c^{\ast}}\left[-\frac{\omega+d_{31}}{D(\omega)}
-\frac{\left(\omega+d_{21}\right)}{\kappa_{12}}\left(\frac{1}{V_{{\rm
g}}}-\frac{1}{c}\right)\right],\\
&& A_{11}^{(2)}=\frac{[i\Gamma_{23}-2|\Omega_c|^2 M]N -i\Gamma_{12}\left(\frac{|\Omega_c|^2}{D(\omega)^{\ast}d_{32}^{\ast}}
-\frac{|\Omega_c|^2}{D(\omega)d_{32}}\right)}{-\Gamma_{12}\Gamma_{23}-
i\Gamma_{12}|\Omega_c|^2 M},\\
&& A_{33}^{(2)}=\frac{1}{i\Gamma_{12}}\left(N-i\Gamma_{12}A_{11}^{(2)}
\right)|F|^2e^{-2\bar{\alpha}z_2},\\ &&A_{32}^{(2)}=\frac{1}{d_{32}}\left(-\frac{\Omega_c}{D(\omega)}
+2\Omega_cA_{33}^{(2)}+\Omega_cA_{11}^{(2)}\right),
\end{eqnarray}
\end{subequations}
where
$\bar{\alpha}=\epsilon^{-2}\alpha=\epsilon^{-2}$Im$[K(\omega)]$,
$M=1/d_{32}-1/d_{32}^{\ast}$,
and $N=(\omega+d_{31}^{\ast})/D(\omega)^{\ast}-(\omega+d_{31})/D(\omega)$.

With the above result we proceed to third order. The divergence-free condition
in this order yields the nonlinear equation for $F$:
\begin{equation} \label{NLS}
i\frac{\partial F}{\partial
z_2}-\frac{1}{2}K_2\frac{\partial^2F}{\partial
t_1^2}-We^{-2\bar{\alpha}z_2}|F|^2F=0,
\end{equation}
where $K_2\equiv \partial^{2} K/\partial \omega^{2}$ and
$W=-\kappa_{12}\left[\Omega_c^{\ast}A_{32}^{(2)}
+(\omega+d_{31})(2a_{11}^{(2)}+A_{33}^{(2)})\right]/D(\omega)$
are dispersion and nonlinear (Kerr) coefficients, respectively.

\subsection{Ultraslow optical solitons}

Combining Eqs.~(\ref{EQ1}) and (\ref{NLS}) and returning to the original variables
we obtain
\begin{equation}\label{NLSa}
i\left(\frac{\partial}{\partial z}+\alpha\right)U-\frac{K_2}{2}\frac{\partial^2U}{\partial\tau^2}-W|U|^2U=0,
\end{equation}
where $\tau=t-z/V_g$ and $U=\epsilon Fe^{-\bar{\alpha}z_2}$. Due to the resonant character of the system,
the NLS equation (\ref{NLSa}) has complex coefficients.
Generally, such equation does not allow soliton solution. However, if the imaginary part of the
coefficients can be made much smaller than their real part, it is possible to form solitons in the
system. We shall show below this can indeed  be achieved in the present EIT system.

For the ultracold Rydberg atoms with the energy-levels assigned by (\ref{param1}) and the system parameters given by (\ref{param2}), we obtain $ K_0=(0.23+i0.0002)$ ${\rm cm^{-1}}$, $ K_1=(1.15+i0.0009)\times10^{-7}$ ${\rm cm^{-1}s}$, $ K_2=(1.82+i0.05)\times10^{-15}$ ${\rm cm^{-1}s^2}$, $ W=(2.59+i0.002)\times10^{-18}$ ${\rm cm^{-1}s^2}$ when selecting $\Delta_2= 700$ $ {\rm MHz}$, $\Delta_3=2$ $ {\rm MHz}$, $\kappa_{12}=1\times 10^{10}$ $ {\rm cm^{-1}}{\rm s^{-1}}$, and $\Omega_c=300$ $ { \rm MHz}$.
We see that the imaginary parts of the coefficients of the NLS equation (\ref{NLSa})
are indeed much smaller than their corresponding real parts. As a result,  Eq.~(\ref{NLSa})
can be approximated as the following dimensionless form
\begin{equation}\label{NLSb}
i\frac{\partial u}{\partial s}+\frac{\partial^2u}{\partial\sigma^2}+2u|u|^2=i\nu u,
\end{equation}
with $s=-z/(2L_D)$, $\sigma=\tau/\tau_0$, and $u=U/U_0$, and $\nu=2L_D/L_A$. Here  $L_D\equiv {\tau_0}^2/|\tilde K_2|$, $L_A\equiv 1/(2\alpha)$, and $U_0\equiv (1/\tau_0)\sqrt{|\tilde K_2/\tilde W|}$ are
characteristic dispersion length, absorption length, and Rabi frequency of the probe field, respectively.
Note that  in order to obtain soliton solutions we have assumed $L_D$ is equal to
$L_{\rm NL}\equiv 1/(U_0^2\tilde{W})$ (the characteristic nonlinearity length). The tilde symbol
means taking real part (e.g. $\tilde{K}_2={\rm Re}(K_2)$).

If taking $\tau_0=1.0\times 10^{-7}$ ${\rm s}$, we obtain $U_0=2.65\times 10^{8}$ $ {\rm s^{-1}}$, $L_D=$ $L_{\rm NL}=5.48$ $\rm cm$, $L_A=2500$ $\rm cm$. Because $L_A$ is much larger than $L_D$ and $L_{\rm NL}$ which gives $\nu=0.0044$, the absorption term on the right hand side of Eq.~(\ref{NLSb}) can be neglected in the leading-order approximation. Thus we obtain a standard NLS equation that is completely integrable and allows various soliton solutions. After returning to the original variables,
the half Rabi frequency of the probe
field corresponding to single-soliton solution reads
\begin{equation}\label{Sol}
\Omega_p(z,t)=\frac{1}{\tau_0}\sqrt{\frac{\tilde{K_2}}{\tilde{W}}}{\rm
sech}\left[\frac{1}{\tau_0}\left(t-\frac{z}{\tilde{V_g}}\right)\right]{\rm exp}{\left[i\tilde{K}_0z-i\frac{z}{2L_D}\right]},
\end{equation}
where $\tilde{K}_0=\tilde{K}(\omega)\mid_{\omega=0}$. With the above system parameters, we obtain
\begin{equation}
\tilde V_g=3\times10^{-4}\,c,
\end{equation}
i.e. the soliton has an ultraslow propagating velocity, which is essential for its storage and retrieval
considered in the next section.

\begin{figure}
\includegraphics[scale=0.8]{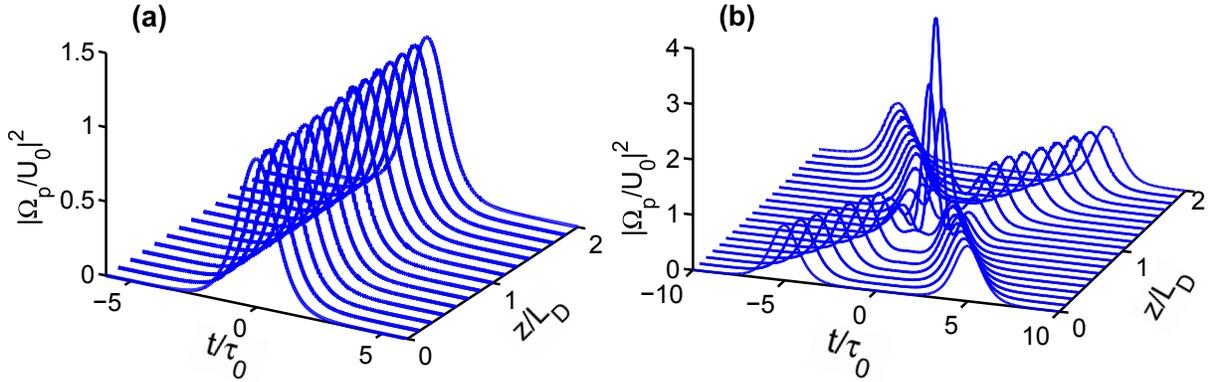}\\
\caption{(Color online) Propagation of the ultraslow optical soliton and the interaction
between two solitons.
(a) Waveshape $|\Omega_p/U_0|^2$ of the ultraslow optical soliton as a function of $z/L_D$ and $t/\tau_0$.  (b) Collision between two ultraslow optical solitons.} \label{collision}
\end{figure}

Shown in Fig.~\ref{collision}(a)
%
%%%%%%%%%%%%%%%%%%%%%%%%%%%%%%%%%%%%%%%%%%%%%%%%%%%%%%%
%%%%%%%%%%%%%%%%%%%%%%%%%%%%%%%%%%%%%%%%%%%%%%%%%%%%%%%%
%
is the numerical result of the waveshape $|\Omega_p/U_0|^2$ of the ultraslow optical soliton as a function of $z/L_D$ and $t/\tau_0$. When making the calculation, Eq.~(\ref{NLSb}) is used with the solution (\ref{Sol}) as an initial condition. Fig.~{\ref{collision}(b) shows the
collision between two ultraslow optical solitons, with the initial condition given by
$u(0,\sigma)={\rm sech}(\sigma-5)\exp(-i\sigma)+{\rm sech}(\sigma+5)\exp (i\sigma)$.
We see that the ultraslow optical solitons are robust during the propagation and the collision.

It is easy to calculate the threshold of the optical power density $P_{\rm max}$ for generating the ultracold optical soliton predicted above by using Poynting's vector~\cite{huang2005}. We obtain
\begin{equation}
P_{\rm max}=9.38\times10^{-5}\,\,{\rm W}.
\end{equation}
Thus, to generate the ultraslow optical solitons in the system, very low input power is needed.

\section{Storage and retrieval of the ultraslow optical solitons}\label{sec4}

In a pioneered work~\cite{Fleischhauer2000},  Fleischhauer and Lukin showed the possibility of
storage and retrieval of optical pulses in a three-level atomic system with a $\Lambda$-type level
configuration. They demonstrated that, when switching on control field, probe pulse propagates
in the atomic medium with nearly vanishing absorption; by slowly switching off the control field
the probe pulse disappears and gets stored in the form of atomic coherence; when the control field
is switched on again the probe pulse reappears. However, the intensity of the probe pulse used in
Ref.~\cite{Fleischhauer2000} and a series of studies carried out later on (see Refs.~\cite{LDBH,phillip2001,Longd,schnorr2009,Zhao,zhang2009,Rad,Dudin10,
Yangf,PFMWL,dudin2013,Dai,heinze2013} and references therein)
is weak, i.e. systems used in those studies~\cite{Fleischhauer2000,LDBH,phillip2001,Longd,schnorr2009,Zhao,zhang2009,Rad,Dudin10,
Yangf,PFMWL,dudin2013,Dai,heinze2013} work in linear regime. Now, we extend these studies
into a weakly nonlinear regime, and demonstrate that it is possible to realize the storage and retrieval of
ultraslow optical solitons in the ladder-type atomic system via EIT.

To this end, we consider the solution of the MB equations presented in Sec.~\ref{Sec2}. We stress
that for the storage and retrieval of optical solitons the dynamics of the control-field must be taken into account, i.e. Eq.~(\ref{Max12}) must be solved together with
Eqs.~(\ref{bloch}) and (\ref{Max11}). Because in this case analytical solutions are not available, we
resort to numerical simulation.

Fig.~\ref{st} shows
%
%%%%%%%%%%%%%%%%%%%%%%%%%%%%%%%%%%%%%%%%%%%%%%%%%%%%%%%
\begin{figure}
\includegraphics[scale=0.8]{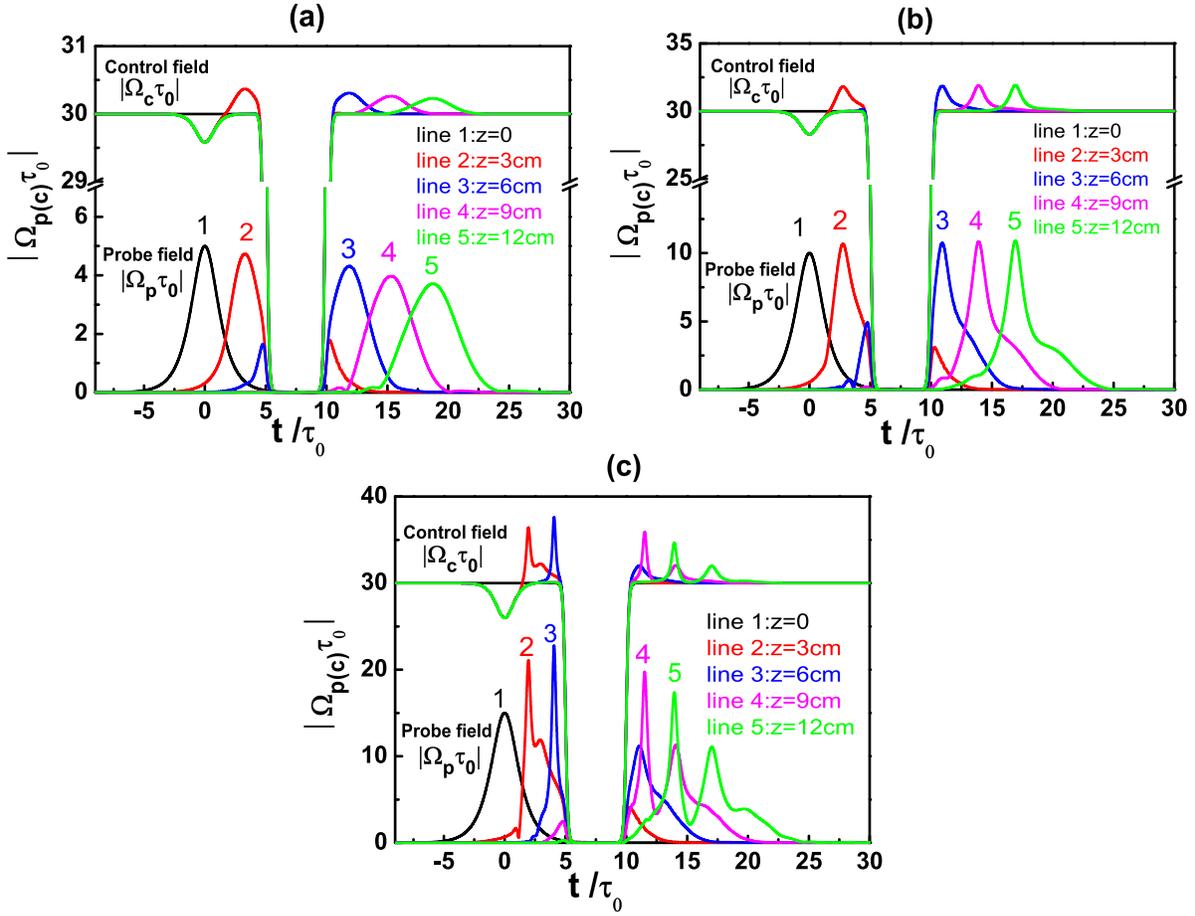}\\
\caption{(Color online) Time evolution of $|\Omega_p\tau_0|$ and $|\Omega_c\tau_0|$
as functions of $z$ and $t$ for different input light intensities.
(a) Storage and retrieval of weak (i.e. linear) pulse,
with $\Omega_p(0,t)\tau_0=5\,{\rm sech}(t/\tau_0)$. (b)
Storage and retrieval of soliton pulse, with $\Omega_p(0,t)\tau_0=10\,{\rm
sech}(t/\tau_0)$. (c) Storage and retrieval of strong pulse, with $\Omega_p(0,t)\tau_0=15\,{\rm
sech}(t/\tau_0)$. The lines from 1 to 5 in each panel correspond to
$z$=0, 3 cm, 6 cm, 9 cm, and 12 cm, respectively.} \label{st}
\end{figure}
%%%%%%%%%%%%%%%%%%%%%%%%%%%%%%%%%%%%%%%%%%%%%%%%%%%%%%%%
%
the time evolution of $|\Omega_p\tau_0|$ and $|\Omega_c\tau_0|$
as functions of $z$ and $t$ for different input light intensities.
In the simulation, the switching-on and the switching-off the control field is
modeled by the combination of two hyperbolic tangent functions with the form
\begin{equation}
\Omega_c(0,t)=\Omega_{c0}\left\{1-\frac{1}{2}{\rm tanh}\left[\frac{t-T_{\rm
off}}{T_s}\right]+\frac{1}{2} {\rm tanh}\left[\frac{t-T_{\rm on}}{T_s}\right]\right\},
\end{equation}
where $T_{\rm off}$ and $T_{\rm on}$ are respectively the times of
switching-off and the switching-on of the control field with a switching time approximately
given by $T_s$. The system parameters are chosen from a typical cold alkali
$^{87}$Rb atomic gas with
$\Gamma_{12}/2\pi=6$ $ {\rm MHz}$, $\Gamma_{23}/2\pi=3.2$ $ {\rm KHz}$, $\gamma_{21}\tau_0\approx1.88$, $\gamma_{31}\tau_0\approx10^{-4}$, $\Delta_2\tau_0=70$, $\Delta_3\tau_0=0.2$, $\kappa_{12}\tau_0=1\times10^{3}$ ${\rm cm^{-1}}$, $\kappa_{23}\tau_0=2\times10^{3}$ ${\rm cm^{-1}}$, $\Omega_{c0}\tau_0=30$,
$\rm T_s/\tau_0=0.2$, $T_{\rm off}/\tau_0=5$, $T_{\rm on}/\tau_0=10$, with $\tau_0=10^{-7}{\rm s}$.
The waveshape of the input probe pulse is taken as a hyperbolic secant one, i.e., $\Omega_p(0,t)=\Omega_{p0}\, {\rm sech}(t/\tau_0)$, with different $\Omega_{p0}$ to represent weak (i.e. linear), soliton (i.e. weak nonlinear), and strong probe regimes. Lines from 1 to 5 are for $z=0$, 3 cm, 6 cm, 9 cm, and 12 cm, respectively.

Shown in Fig.~\ref{st}(a) is the result for a weak (i.e. linear) probe pulse, where $\Omega_p(0,t)\tau_0=5\,{\rm sech}(t/\tau_0)$. In this case, the system is dispersion-dominant. Storage and retrieval of light pulses are possible, but the probe pulse broadens fast before and after the storage, which is not desirable for practical applications because light information will be lost after the storage.

Fig.~\ref{st}(b) shows the result for a weak nonlinear (i.e. soliton) probe pulse,
where $\Omega_p(0,t)\tau_0=10\,{\rm sech}(t/\tau_0)$. In this situation, the system works in the regime with a balance between dispersion and nonlinearity. We see that the probe pulse evolves firstly into a soliton
(i.e. its pulse width is narrowed) before the storage; later on the soliton is stored in the atomic medium (i.e. $\Omega_p=0$ when $\Omega_c$ is switched off); then the soliton is retrieved  after the storage (when $\Omega_c$ is switched on). The retrieved soliton has nearly the same waveshape as that before the storage.

Shown in Fig.~\ref{st}(c) is the result for a strong probe pulse, where $\Omega_p(0,t)\tau_0=15\,{\rm sech}(t/\tau_0)$. In this case, the system works in a strong nonlinear (i.e. nonlinearity-dominant)  regime
and hence a stable soliton is not possible. From the figure we see that the probe pulse has a significant distortion, especially some new peaks are generated. Due to the large distortion, the light information will be lost fast even before the storage.

From the result of Fig.~\ref{st}, we conclude that, comparing with the linear pulse and the strong nonlinear pulse, the soliton pulse is desirable for the storage and retrieval. One may ask the question how the optical soliton is stored into the atoms when both the probe and control fields have vanishing value. In fact, during the light storage the probe-field energy is converted into atomic degrees of freedom, i.e. the atomic coherence $\sigma_{13}$ has non-vanishing value even when both $\Omega_c$ and $\Omega_p$ are zero.

Shown in Fig.~\ref{coherence}
%
%%%%%%%%%%%%%%%%%%%%%%%%%%%%%%%%%%%%%%%%%%%%%%%%%%%%%%%
\begin{figure}
\includegraphics[scale=0.8] {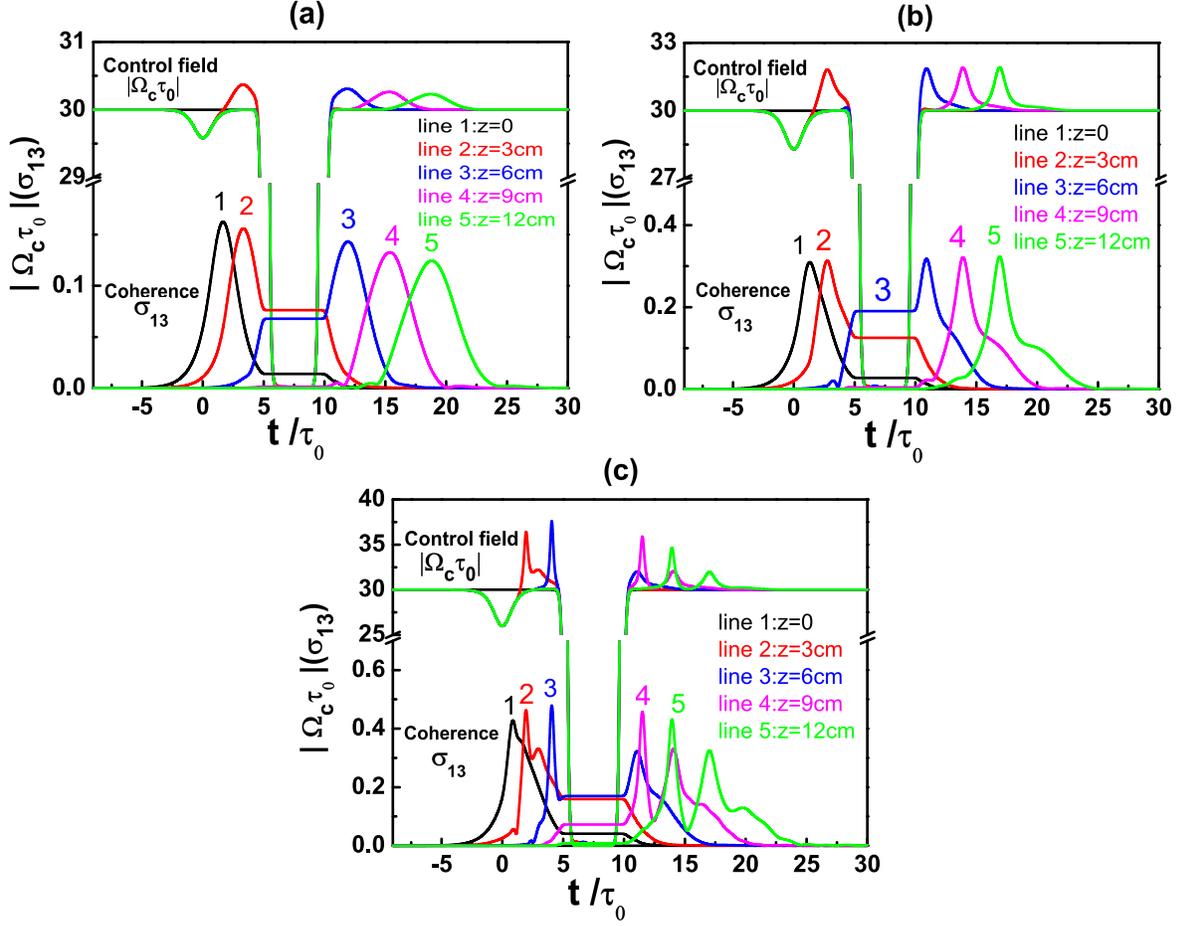}\\
\caption{(Color online) The atomic coherence $\sigma_{13}$ as
functions of distance $z$ and time $t$ for different input light intensities.
Corresponding evolution of $|\Omega_c\tau_0|$ is also shown.
Initial probe pulses used in
panels (a),(b), and (c) are the same as those used in Fig.~3(a), Fig.~3(b), and Fig.~3(c), respectively.
The lines from 1 to 5 in each panel correspond to
$z$=0, 3 cm, 6 cm, 9 cm, and 12 cm, respectively.
}\label{coherence}
\end{figure}
%%%%%%%%%%%%%%%%%%%%%%%%%%%%%%%%%%%%%%%%%%%%%%%%%%%%%%%%
%
is the result of $\sigma_{13}$ for different input light intensities as functions of $z$ and $t$.
Corresponding evolution of $|\Omega_c\tau_0|$ is also plotted.
Initial probe pulses used in each panels of the figure are the same as those used in Fig.~\ref{st}.
The lines from 1 to 5 in each panel of the figure correspond to $z$=0, 3 cm, 6 cm, 9 cm, and 12 cm, respectively. From Fig.~\ref{coherence} combining with Fig.~\ref{st}
we see that indeed $\sigma_{13}\neq 0$ in the time interval when $\Omega_c=\Omega_p=0$.
Since the probe pulse is stored in the form of atomic coherence $\sigma_{13}$ when the control field is switched off and is retained until the control field is switched on again, the atomic coherence $\sigma_{13}$ can be taken  as the intermediary for the storage and retrieval of the probe pulse.

Now we give a simple explanation on the numerical result given above. Note that
when the control pulse is switching off, the probe pulse becomes nearly zero. Thus
in the weak nonlinear regime, the probe pulse can be approximated as
\begin{equation}\label{solnew}
\Omega_p(z,t)\approx \left\{
\begin{array}{lll}
\frac{A}{\tau_0}\sqrt{\frac{\tilde{K}_2}{\tilde{W}}}{\rm sech}\left[\frac{1}{\tau_0}\left(t-\frac{z}{\tilde{V_g}}\right)\right]e^{i[\tilde{K}_0-1/(2L_D)]z},\,\,   {\rm for}\,\, t<{\rm T_{off}},\\
0,\,\,\,\,   {\rm for}\,\, {\rm T_{off}}\leq t\leq {\rm T_{on}},\\
\frac{B}{\tau_0}\sqrt{\frac{\tilde{K}_2}{\tilde{W}}}{\rm sech}\left[\frac{1}{\tau_0}\left(t-\frac{z}{\tilde{V_g}}\right)\right]e^{i[\tilde{K}_0-1/(2L_D)]z+i\phi_0},\,\,   {\rm for}\,\, t>{\rm T_{on}},
\end{array}
\right.
\end{equation}
where $A$ and $B$ are constants depending on initial condition, $\phi_0$ is a constant phase factor.

From both Fig.~\ref{st} and Fig.~\ref{coherence}, we see that the control field is changed before and after  the storage of the probe field. To analyze the dynamics (depletion) of the control field before and after the storage of the probe soliton, we solve Eq.~(\ref{Max12}) using a perturbation method.
The numerical result shown in Fig.~\ref{st} and Fig.~\ref{coherence} suggests us to
make the perturbation expansion
\begin{equation} \label{control}
\Omega_{c}=\Omega_{c}^{(0)}+\epsilon\Omega_{c}^{(1)}+\epsilon^2\Omega_{c}^{(2)},
\end{equation}
which is valid for the time interval before and after the probe soliton storage where the leading-order
of $\Omega_c$ (i.e. $\Omega_c^{(0)}$) has a large value.
Substituting the expansion (\ref{control}) into Eq.~(\ref{Max12}) and solving the equations for $\Omega_{c}^{(l)}$ ($l=0,1,2$),
we obtain the following conclusions: (i) $\Omega_{c}^{(0)}$ is a constant,
which corresponds to the horizontal line in the upper part of Fig.~\ref{st} and Fig.~\ref{coherence}.
(ii) $\Omega_{c}^{(1)}(t,z)=\Omega_{c}^{(1)}(t-z/c)$ describes a hole  below the horizontal line
in the upper part of Fig.~\ref{st} and Fig.~\ref{coherence}, which
propagates with velocity $c$ (i.e. the light speed in vacuum). The concrete form of
$\Omega_{c}^{(1)}$ relies on initial condition.
(iii) $\Omega_c^{(2)}$ satisfies the equation $i\partial \Omega_{c}^{(2)}/\partial z=-\kappa_{23}\sigma_{32}^{(2)}$.
Thus we obtain
\begin{equation}\label{Dep}
\Omega_{c}^{(2)}=i\frac{\kappa_{23}
\Delta_{3}}{(|\Omega_c|^2-\Delta_{2}\Delta_{3})^2}\frac{1}{\tau_0^2}
\frac{\tilde{K}_2}{\tilde{W}}\tanh\left[\frac{1}{\tau_0}\left(t-\frac{z}{\tilde{V}_g}\right)\right],
\end{equation}
which has propagating velocity $\tilde{V_g}$ and contributes a small hump to the horizontal line in the upper part of Fig.~\ref{st} and Fig.~\ref{coherence}. That is to say, the hump propagates with the same velocity as the probe soliton. Physically, the appearance of the control field hump (depletion) is due to the energy exchange between the control field and the probe field via the atomic system as an intermediary.

In addition, we can also provide a simple theoretical explanation on the behavior observed in Fig.~\ref{coherence}(b), where, before and after the storage of the probe soliton, $\sigma_{13}$ behaves
like a soliton but it is constant during the storage. First, let us consider the time interval before and after the storage of the probe soliton where $\Omega_c\approx \Omega_{c}^{(0)}$ (for simplicity we neglect the small hump described by Eq.~(\ref{Dep})). In this region, the perturbation expansion given in
Sec.~\ref{sec3}A is still valid. Thus the result obtained there can be used here. From
Eqs.~(\ref{FO1}) and (\ref{FO3}), when evaluated at the center frequency of the probe pulse
(i.e. $\omega=0$) we obtain
\begin{equation}
\sigma_{13}=-\frac{\Omega_c^*}{D^*(0)}\Omega_p^*\approx -\frac{\Omega_c^{*(0)}}{D^*(0)}
\frac{1}{\tau_0}\sqrt{\frac{\tilde{K_2}}{\tilde{W}}}{\rm
sech}\left[\frac{1}{\tau_0}\left(t-\frac{z}{\tilde{V_g}}\right)\right]{\rm exp}{\left[-i\tilde{K}_0z+i\frac{z}{2L_D}\right]}.
\end{equation}
We see that, before and after the storage of the probe soliton, $\sigma_{13}$ is also soliton with the propagating velocity $\tilde{V_g}$, as expected.

The behavior of $\sigma_{13}$ in the time interval during the probe-soliton
storage can not be explained by using the perturbation theory developed in Sec.~\ref{sec3}A because
in this case $\Omega_c$ is a small quantity. To solve this problem, we start to consider
the Bloch Eq.~(\ref{bloch}) directly~\cite{milo}. Since $d_{31}\sigma_{31}$ and $\Omega_p\sigma_{32}$ are small,
by Eq.~(\ref{bloch5}) we have
\begin{equation}\label{si21}
\sigma_{21}\approx -\frac{i}{\Omega_c}\frac{\partial\sigma_{31}}{\partial t}.
\end{equation}
Furthermore, since $\sigma_{11}\approx 1$ and $\sigma_{22}\approx 0$, Eq.~(\ref{bloch4}) gives
\begin{equation}\label{si31}
\sigma_{31}\approx -\frac{\Omega_p}{\Omega_c^*}+\frac{1}{i\Omega_c^*}
\left( \frac{\partial}{\partial t}-id_{21}\right) \sigma_{21}.
\end{equation}
Substituting Eq.~(\ref{si21}) into Eq.~(\ref{si31}) we obtain
\begin{eqnarray}\label{si311}
\sigma_{13}& & = -\frac{\Omega_p^*}{\Omega_c}-\frac{1}{|\Omega_c|^2}
                       \left( \frac{\partial}{\partial t}+id_{21}\right) \sigma_{13}\nonumber\\
           & & \approx -\frac{\Omega_p^*}{\Omega_c}.
\end{eqnarray}
Although in the time interval of the storage both the probe and control fields tend into zero,
the ratio $\Omega_p^*/\Omega_c$ can keep a finite constant value, as shown  in the
numerical simulation presented in Fig.~\ref{coherence}. The physical reason is that
in our study the system starts from the dark state $|D\rangle=\Omega_c^*|1\rangle-\Omega_p|3\rangle
=\Omega_c^*(|1\rangle-(\Omega_p^*/\Omega_c)^*|3\rangle)$ and it approximately keeps in this dark state during time
evolution. As a result, the atomic coherence $\sigma_{13}$ can have a nonzero value even
both $\Omega_c$ and $\Omega_p$ are small.

\section{SUMMARY}\label{sec5}

In the present contribution, we have proposed a method for obtaining stable nonlinear optical pulses
and realizing  their storage and retrieval
in an ultracold ladder-type three-level atomic gas via EIT.
Starting from the MB equations, we have derived a NLS equation
governing the evolution of probe-field envelope. We have shown that
optical solitons with ultraslow propagating velocity and extremely low
generation power can be created in the system. Furthermore,
we have demonstrated that such ultraslowly propagating, ultralow-light level optical solitons
can be stored and retrieved by switching off and on the control field.
Because of the balance between dispersion and nonlinearity,
the ultraslow optical solitons are robust during propagation,
and hence their storage and retrieval are more desirable than that of
linear optical pulses. Our study provides a possibility of realizing
light information storage and retrieval by using solitonlike nonlinear optical pulses.

\acknowledgments This work was supported by the NSF-China under Grant Numbers 11174080 and 11105052.

%%%%%%%%%%%%%%%%%%%%%%%%%%%%%%%%%%%%%%%%%%%%%%%%%%%%%%%%%%%%%%%%%%%%%%%%%%%%%%%%%%%%%%%%%%%


\begin{references}
\bibitem{note0}The frequency and wave number of the probe field are given by
$\omega_p+\omega$ and $k_p+K(\omega),$ respectively. Thus
$\omega=0$ corresponds to the center frequency of the probe
field.
\end{references}
\end{document}